\documentclass[]{aa}

\newcommand{\apropto}{\;
  \raise0.3ex\hbox{$\propto$\kern-0.75em\raise-1.1ex\hbox{$\sim$
  }}\;\hskip-2pt }
\usepackage[dvips]{graphicx}

\newcommand{\lta}{\;
  \raise0.3ex\hbox{$<$\kern-0.75em\raise-1.1ex\hbox{$\sim$
  }}\;\hskip-2pt }
\newcommand{\gta}{\;
  \raise0.3ex\hbox{$>$\kern-0.75em\raise-1.1ex\hbox{$\sim$
  }}\;\hskip-2pt }

\begin{document}
\title{Ram pressure effects in the galactic plane and galactic dynamos in the no-$z$ approximation
}

   \author{D.~Moss\inst{1} \and D.~Sokoloff\inst{2} \and R.~Beck\inst{3}}

   \offprints{D.Moss}

   \institute{School of Mathematics, University of Manchester, Oxford Road,
Manchester, M13 9PL, UK \and Department of Physics, Moscow
University, 119992 Moscow, Russia \and MPI f\"ur Radioastronomie,
Auf dem H\"ugel 69, 53121 Bonn, Germany}

   \date{Received ..... ; accepted .....}

\abstract{The magnetic field of galaxies is believed to be produced
by internal dynamo action, but can be affected by motion of the
galaxy through the surrounding medium. Observations of polarized
radio emission of galaxies located in galaxy clusters have revealed
noticeable features of large-scale magnetic configurations,
including displacements of the magnetic structures from the optical
images and tails, which are possible imprints of ram pressure
effects arising from motion of the galaxies through the intracluster
medium.} {We present a quantitative dynamo model which attempts to
describe the above effects. In contrast to the traditional problem
of a wind affecting a body with a prescribed magnetic field, we
investigate how a non-magnetized wind flow affects a magnetic field
that is being self-excited by galactic dynamo action.} {In order to
isolate the leading physical effects we exploit a simple dynamo
model that can describe relevant effects. In particular, we use what
is known as the 'no-$z$' approximation for the mean-field dynamo
equations.}{In a suitable parametric range we obtain displacements
of the large-scale magnetic field, as well as magnetic tails.
However, the specific details of their locations are quite
counterintuitive. The direction of displacement is perpendicular to,
rather than parallel to, the wind direction. The point at which the
tail emerges from the galaxy depends on details of the model. The
tail is eventually directed downstream. In the simplest case the
magnetic tail begins in the region where the wind decreases the
total gas velocity. Any wind that penetrates the galaxy modifies the
intrinsic dynamo action. These features are different from those found in
ram-pressure models.} {Any determination of galactic motion through
the cluster medium from observational data needs to take the effects
of dynamo action into account.}


\titlerunning{Ram pressure effects and galactic dynamos}
\authorrunning{Moss et al.}

\maketitle

\section{Introduction}

The formation and subsequent evolution of large-scale galactic
magnetic fields are believed to be a result of galactic physical
processes, mainly induction effects of differential rotation and
mirror-asymmetric galactic turbulence (e.g. Beck et al. 1996).
Galaxies are surrounded by the intergalactic medium and participate
in various intergalactic interactions which somehow modify the
``intrinsic'' magnetic properties of galaxies. The corresponding
effects on galactic magnetic structures naturally attract interest
(e.g. Williams et al. 2002). A straightforward example of this is
provided by the propagation of galaxies through the medium pervading
galaxy clusters. It appears quite natural to attempt to attribute
the existence of various galactic tails to the effects of such
propagation. Such a morphological interpretation {\it a priori}
appears natural; however it requires verification by modelling,
including effects of galactic dynamo action.

Contemporary radio polarimetric observations provide several
examples of large-scale galactic magnetic configurations which
appear possibly to be affected by such propagation effects. In
particular, Vollmer et al. (2007) and We\.zgowiec et al. (2007)
observed magnetic fields of large spiral galaxies in the Virgo
Cluster in order to study interactions of galaxies with the cluster
environment. (Further relevant results can be found in Kantharia et
al. 2008; Vollmer et al. 2010; We\.zgowiec et al. 2011, see also
Yoshida et al. 2012.) A straightforward intention of this work was
to use the proximity of the Virgo Cluster to isolate effects of the
high-velocity tidal interactions and the effects of ram pressure
stripping by the intracluster gas.

\begin{figure}
\centering
\includegraphics[width=0.3\textwidth]{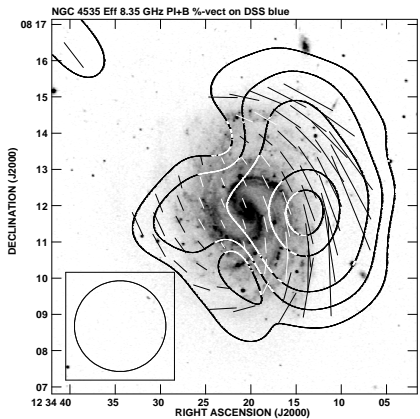}
\caption{Polarized radio emission (contours) and B--vectors of the
spiral galaxy NGC~4535 in the Virgo Cluster, observed at 3.6\,cm
wavelength with the Effelsberg 100-m telescope, smoothed to
3\arcmin\ resolution and overlaid onto a DSS optical image (kindly
provided by Marek We\.zgowiec).} \label{fig:n4535}
\end{figure}

\begin{figure}
\centering
\includegraphics[width=0.35\textwidth]{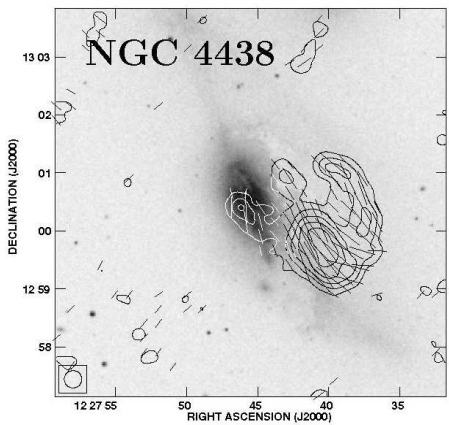}
\caption{Polarized radio emission (contours) and B--vectors of the
spiral galaxy NGC~4535 in the Virgo Cluster, observed at 6.3\,cm
wavelength with the VLA at 18\arcsec\ resolution and overlaid onto a
DSS optical image (from Vollmer et al. 2007).} \label{fig:n4438}
\end{figure}

Indeed, the observational results presented by We\.zgowiec et al.
(2007) demonstrate a magnetic configuration that impressively mimics
the naive expectation of the effects of ram pressure arising from
galactic motion through the surrounding intracluster gas. In
particular, the distribution of total radio intensity in NGC~4535 at
6.3\,cm wavelength (Fig.~5 in the above cited paper) practically
coincides with the optical image while the measures of polarized
emission (Fig.~6 of that paper) are markedly displaced from the
optical image towards the west. This displacement has been confirmed
by observations at 3.6\,cm wavelength (We\.zgowiec et al. 2012) and
hence cannot be caused by Faraday depolarization. Possible
interpretations of this displacement include: (1) the spiral
magnetic field generated in NGC~4535 by a galactic dynamo is
compressed on the western side by ram pressure, (2) weakening of the
galactic dynamo on the eastern side by ram pressure, (3)
amplification and ordering of the magnetic field by shear in the HI
envelope of the galaxy (Vollmer et al. 2010). Note that NGC~4535 is
observed almost face-on, so that if the asymmetry in PI is due to
ram pressure effects then its motion through the ICM is expected to
be at a small or moderate angle with respect to the galactic plane;
in this case the direction of motion can hardly be inferred from
spectroscopic data which gives radial velocities.

We present as an instructive example a map of polarized intensity
and magnetic vectors of NGC~4535 overlaid on the optical image of
this galaxy (Fig.~\ref{fig:n4535}). Note that the maximum of
polarized intensity is located at the periphery of the optical image
while the magnetic field vectors cover the entire optical image.

Another example comes from the observations of NGC~4501. (Fig.~3 in
We\.zgowiec et al. (2007) shows the total intensity and Fig.~4 the
polarized intensity.) Again, the total intensity practically
coincides with the optical image while the polarized emission is
displaced from the optical image. The displacement is less
pronounced than for NGC~4535. NGC~4501 is more inclined towards the
line of sight than NGC~4535, which makes the interpretation less
straightforward. Displacements towards one side of the disc are also
detected in other galaxies observed edge-on, e.g. in NGC~4402
(Vollmer et al. 2007) and in NGC~4388 (We\.zgowiec et al. 2012). On
the other hand, several face-on galaxies of the Virgo Cluster with
large-scale spiral patterns of their magnetic fields, like NGC~4303
and NGC~4321, show no displacement at all, presumably indicating
that they move ``head-on'' through the cluster medium, i.e. almost
along the line of sight (We\.zgowiec et al. 2012).

Each galaxy needs a specific model of gas stripping, gas back-flow
and the effects of the flow on the magnetic field, and these can explain
the observations in many cases (e.g. Vollmer et al. 2009, 2012).
However, these models consider only magnetic fields which are
sheared and amplified by the distorted velocity field of the gas and
neglect the ongoing dynamo action in the galaxy. Not surprisingly,
several features of the magnetic field patterns, such as a
large-scale asymmetry out of the disc plane as in NGC~4192
(We\.zgowiec et al. 2012) or a long magnetic tail in NGC~4438
(Fig.~\ref{fig:n4438}) remained unexplained in these papers.

An interpretation of the displacement as an effect of the ram
pressure interacting with galactic dynamo action needs quantitative
investigation and verification; this is the aim of this paper.

Our idea is to test the interpretation in the most straightforward
way. We present the effect of ram pressure (in the coordinate system
of the galaxy) as a non-magnetized flow of the intracluster gas
which moves through the galaxy parallel to the galactic plane and
interacts with the dynamo-generated magnetic field. We assume that
this motion does not destroy the main drivers of galactic dynamo,
i.e. differential rotation and helical interstellar turbulence, that
are responsible for the generation of the large-scale magnetic field
and subsequently for the polarized emission. We expect to see that
the gas motion (``wind'') displaces the dynamo-generated magnetic
field so that it is no longer centred on the galaxy.

The effect of a wind on magnetic structures around various celestial
bodies is a classical topic that originated in investigations of
interactions between the solar wind and the Earth's magnetic field.
Obviously, the solar wind has a significant effect on the shape of
the Earth's magnetic field, but not on its generation. We consider
here the problem where a flow directly affects the region of
magnetic field generation. We do not consider the galactic magnetic
field to be prescribed {\it a priori}, but rather investigate how
the generation process is modified by the wind.

\section{The dynamo model}
\label{model}

We assume that the ram pressure effects under investigation are
quite robust and its modelling do not require knowledge of finer
details of hydrodynamics and MHD of a particular galaxy (which are
hardly known for NGC~4535 and NGC~4501). Correspondingly, we choose
the most simple galactic dynamo model and include the effects
of interests.

\begin{figure}
\includegraphics[height=0.335\textwidth]{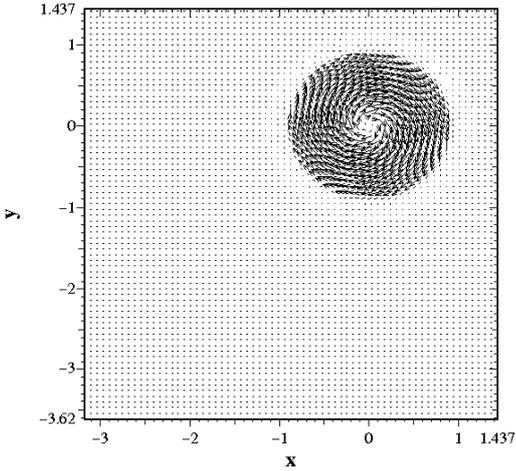}
\caption{Magnetic field vectors for the steady state field without
wind. $R_\alpha=1$, $R_\omega=20$, $r_3=0.9$. The galaxy is assumed
to rotate counterclockwise in this and all other examples.}
\label{fig:nowind}
\end{figure}

\begin{figure}
(a)\includegraphics[height=0.335\textwidth]{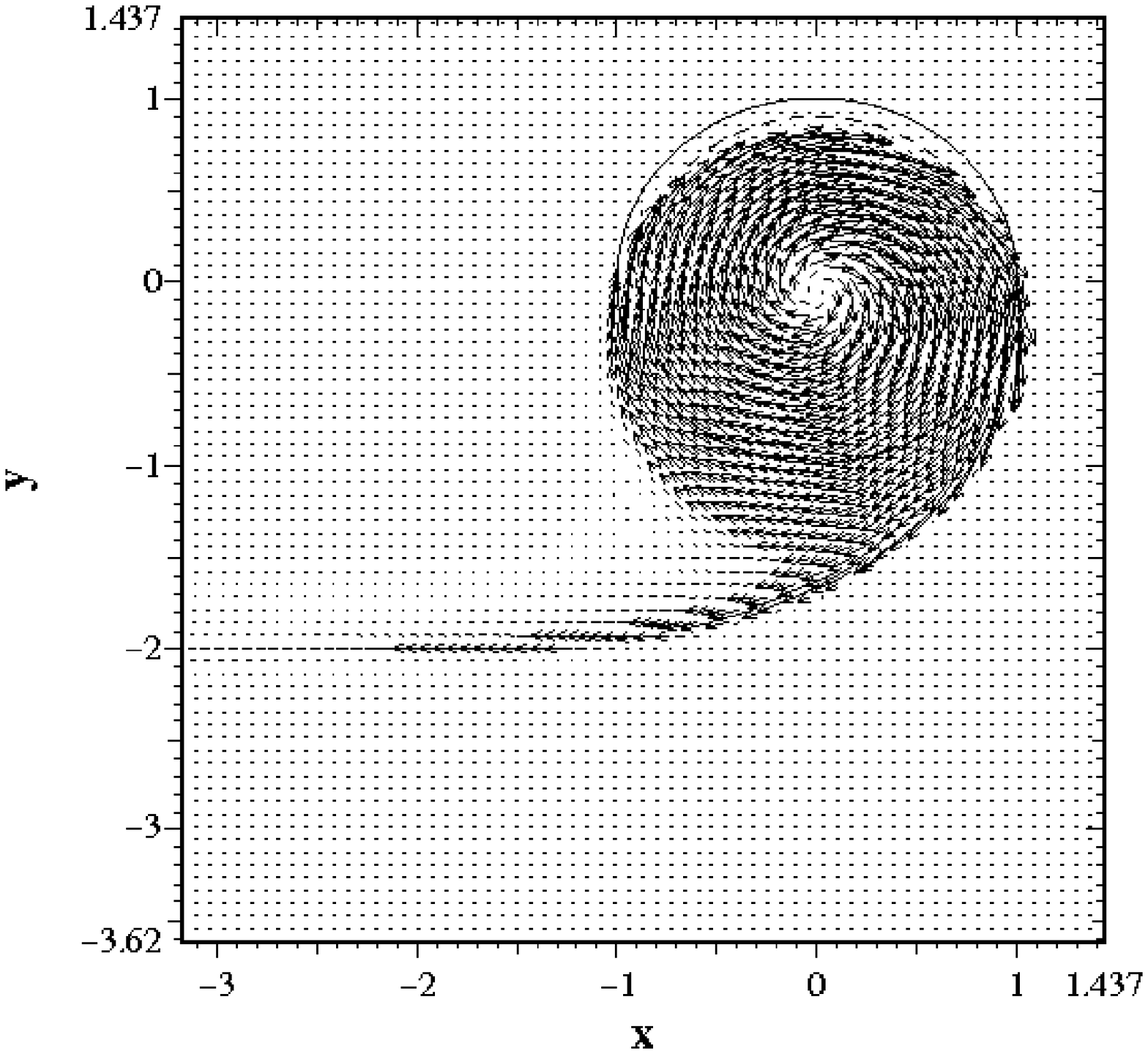}\\
(b)\includegraphics[height=0.335\textwidth]{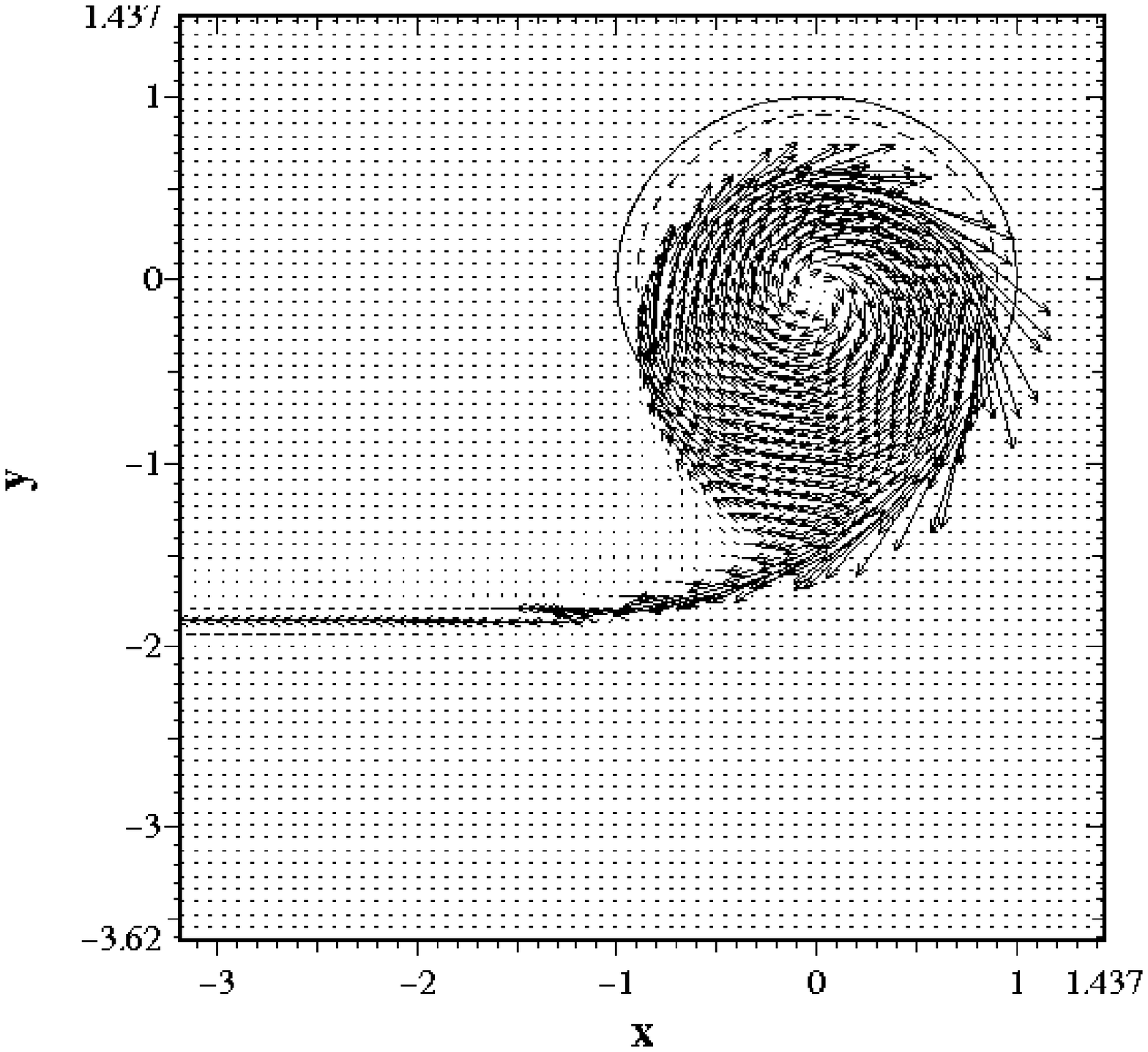}\\
(c)\includegraphics[height=0.335\textwidth]{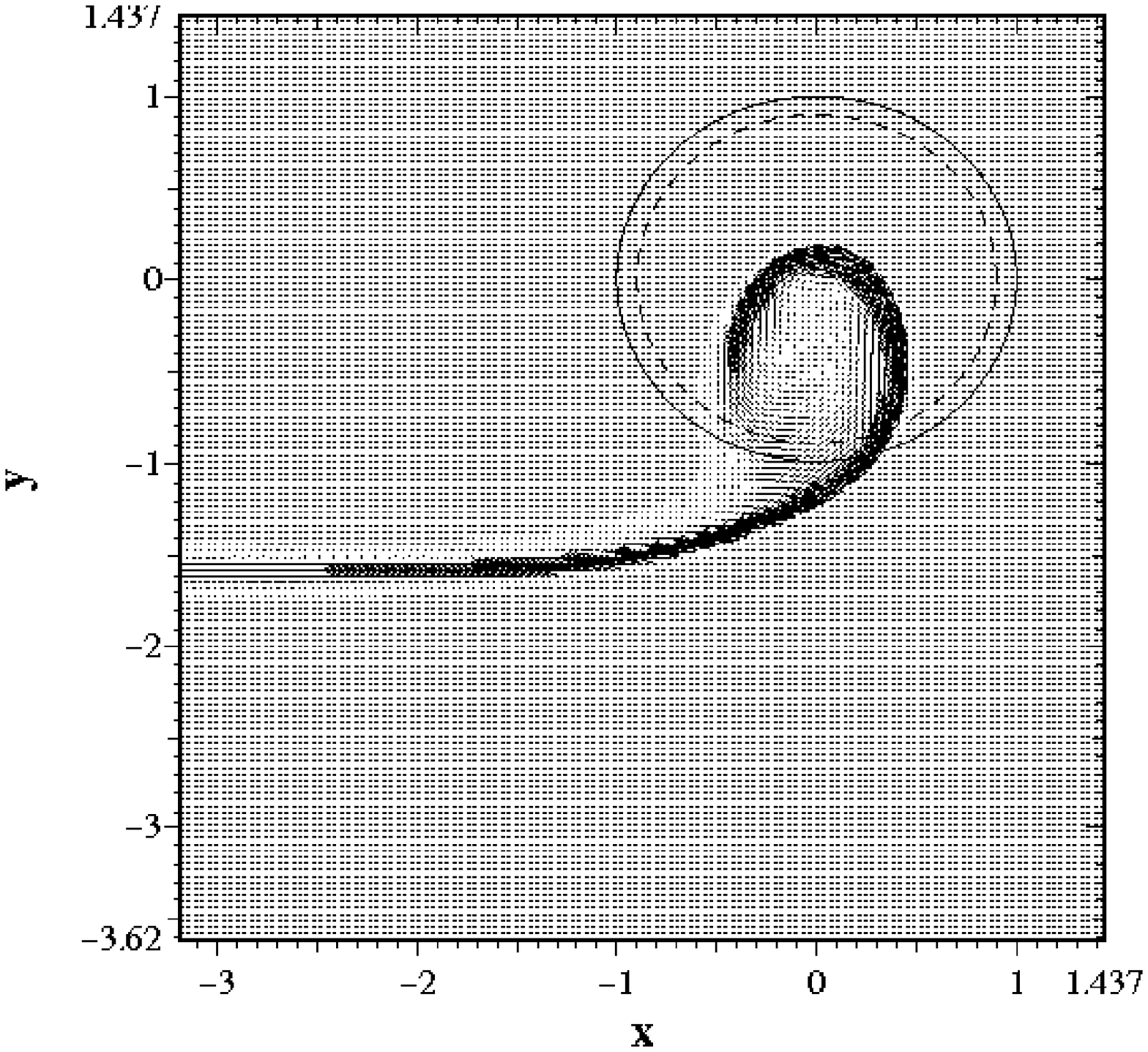}\\
\caption{Vectors of the steady state distribution of magnetic field,
overlaid on circles indicating the location of the dynamo sources
(solid) and $r=r_3$ (broken). (a) $r_3=0.9$, $R_\omega=20, R_{\rm
m}=100$, (b) $r_3=0.9$, $R_\omega=20, R_{\rm m}=150$, (c) $r_3=0.9$,
$R_\omega=10, R_{\rm m}=150$.  The streaming velocity is in the
negative $x$-direction.} \label{fig:wind1}
\end{figure}

We use a version of the 'no-z' model (Subramanian \& Mestel 1993,
Moss 1995), implemented in cartesian coordinates in the
$\alpha\omega$ approximation (see Moss et al. 2012 for more
details). The dynamo-active disc is embedded in an external
diffusive medium, chosen to be large enough that a further increase
in size does not affect the results.

The code solves in the $\alpha\omega$ approximation explicitly for
the field components parallel to the disc plane while the component
perpendicular to this plane (i.e. in the $z$-direction) is given by
the solenoidality condition. The key parameters are the aspect ratio
$\lambda=h/R$, where $h$ corresponds to the semi-thickness of the
warm gas disc and $R$ is its radius, and the dynamo numbers
$R_\alpha=\alpha_0 h/\eta, R_\omega=\Omega_0 h^2/\eta$. $\lambda$
must be a small parameter. $\eta$ is the turbulent diffusivity,
for simplicity assumed uniform throughout the computational region,
and $\alpha_0, \Omega_0$ are typical values of  the
$\alpha$-coefficient and angular velocity respectively. Thus the
dynamo equations become in cylindrical polar coordinates $(r, \phi,
z)$

\begin{eqnarray}
\frac{\partial B_r}{\partial t} & = & -R_\alpha B_\phi-\frac{\pi^2}{4}
B_r \\
\nonumber &&+\,\lambda^2\left(\frac{\partial}{\partial
r}\left[\frac{1}{r}\frac{\partial}{\partial
r}(rB_r)\right]+\frac{1}{r^2}\frac{\partial^2B_r}{\partial\phi^2}-\frac{2}{r^2}\frac{\partial
B_\phi}{\partial\phi}\right) , \label{evolBr}
\end{eqnarray}
\begin{eqnarray}
\frac{\partial B_\phi}{\partial t} & = & R_\omega r
B_r\frac{d\Omega}{dr}-R_\omega\Omega\frac{\partial B_\phi}{\partial
\phi}-\frac{\pi^2}{4} B_\phi \\
 \nonumber
&& +\,\lambda^2\left(\frac{\partial}{\partial
r}\left[\frac{1}{r}\frac{\partial}{\partial r}(rB_\phi)\right]
+\frac{1}{r^2}\frac{B_\phi^2}{\partial
\phi^2}-\frac{2}{r^2}\frac{\partial B_r}{\partial \phi}\right),
\label{evolBphi}
\end{eqnarray}
here $z$ does not appear explicitly. This equation has been
calibrated by introduction of the factors $\pi^2/4$ in the vertical
diffusion terms. These factors arise from a pragmatic examination of
the $z$-dependence of the solutions in the local approximation (e.g.
\cite{ruzetal88}), and their inclusion improves significantly the
agreement between the no-$z$ and local approximations. In principle
in the $\alpha\omega$ approximation the parameters $R_\alpha,
R_\omega$ can be combined into a single dynamo number $D=R_\alpha
R_\omega$, but we choose to keep them separate.

We assume a uniform alpha-effect over the disc region $x^2+y^2\le
1$, and the rotation law in $r<r_3$ is given by
\begin{eqnarray}
&r\frac{d\Omega}{dr}\\
\nonumber & =&\Omega_0\left(-\frac{1}{r R_{\rm gal}}\tanh(\frac{r
R_{\rm gal}}{r_0}) +\frac{1}{r_0\cosh^2(r R_{\rm gal}/r_0)}\right)
\label{rot}
\end{eqnarray}
(see also Moss et al. 2012). This rotation law is truncated at the
fractional radius $r_3\sim 1$, and in $x^2+y^2> r_3^{~2}$ the value
at $r=r_3$ decays smoothly to zero by $r=3$. There is a smooth
transition between the forms in $r<r_3$ and $r>r_3$.

A uniform wind ${\bf U}=-U_0{\bf i}$ blowing in the negative
$x$-direction is imposed. The governing parameter for the wind is
$R_{\rm m}=U_0h/\eta_0$, where $\eta_0$ is the value of the
diffusivity in the disc $x^2+y^2\le 1$. Thus we add to the right
hand sides of the dimensionless Eqs.~(1) and (2) terms corresponding
to the $r$ and $\phi$ components of $R_{\rm
m}\,\lambda\,\nabla\times(-U_0\,{\bf i}\times {\bf B})$ -- see Moss
(1996) for further details.  We note that the treatment of the
``external'' region $r>1$ is subject to increased uncertainty. At
larger galacto-centric distances outside of the ``galaxy'', the
no-$z$ approximation is likely to become more inaccurate in the
absence of a well defined disc. Also the value of the diffusivity in
this region is uncertain -- see also Sect.~\ref{disc}.

\section{Results}
\label{res}

\begin{figure}
\includegraphics[height=0.335\textwidth]{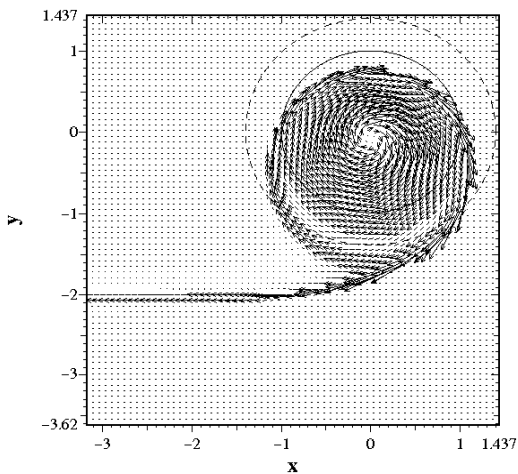}\\
\caption{Vectors of the steady state distribution of magnetic field,
overlaid on circles indicating the location of the dynamo sources
(solid) and $r=r_3$ (broken), $r_3=1.4$, $R_\alpha=1$,
$R_\omega=20, R_{\rm m}=150$.  The streaming velocity is in the
negative $x$-direction.} \label{fig:wind2}
\end{figure}

\begin{figure}
\includegraphics[height=0.365\textwidth]{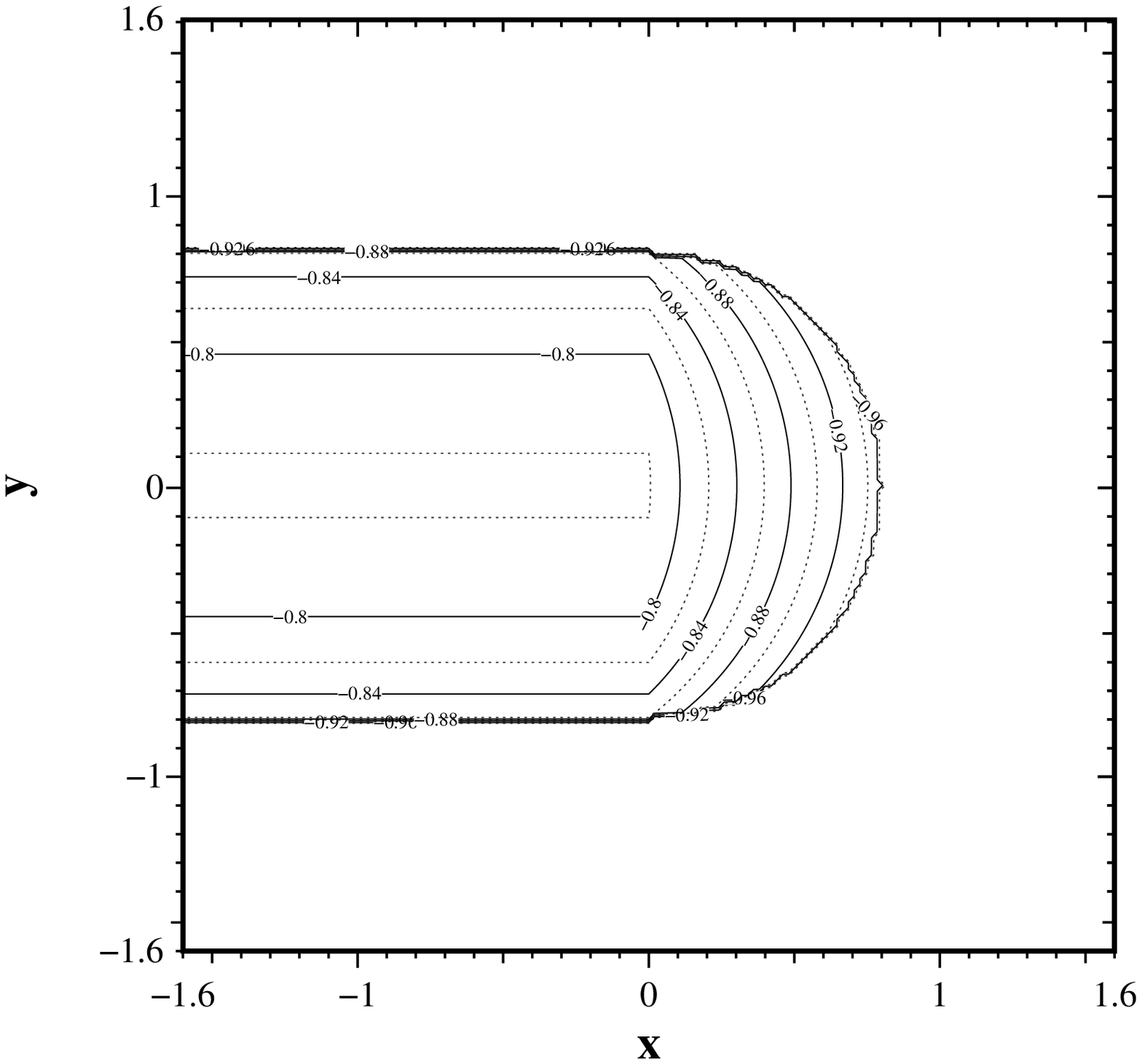}\\
\includegraphics[height=0.335\textwidth]{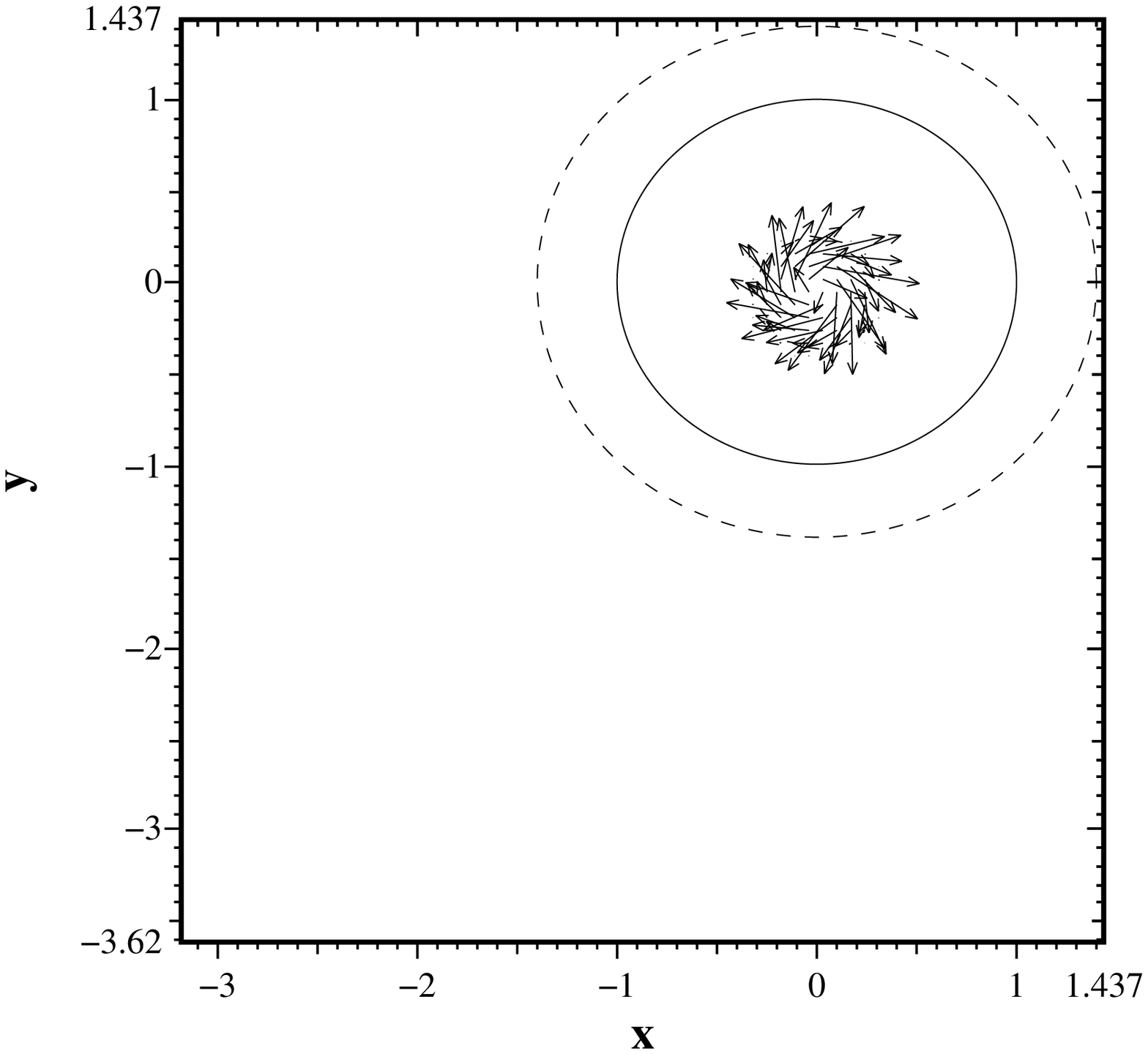}\\
\caption{Model with wind decelerating through the galaxy:
$R_\alpha=2.0$, $R_\omega=30$, $R_{\rm m}=150$. Upper panel -
contours of velocity field; below - the stationary magnetic
configuration. Note that in the upper panel the centre of the plot
is the centre of the galaxy.} \label{fig:wind3}
\end{figure}

We performed extensive numerical simulations with the model. We
first evolve the model with no wind until an approximately steady
state is reached -- see Fig.~\ref{fig:nowind}. (This normally occurs
by $\tau \approx 3-4$.) The initial conditions are a comparatively
weak ($\lta 1\%$ of equipartition strength) quasi-uniform field
(however the steady saturated field configuration is not sensitive
to the initial field provided that it is relatively weak and
more-or-less homogeneous). We then turn on the wind, and evolve
until a steady state is again reached. (In practice, the details of
the start-up procedure make no difference to the final
configuration.)

Putting $R_\alpha=1$, for two values of $r_3$ (the rotation law
transition radius), we computed models with several values of
$R_\omega$ and $R_{\rm m}$. With $r_3=0.9$ we show in
Fig.~\ref{fig:wind1} the final states of the magnetic field for
$R_\omega=20, R_{\rm m}=100, 150$, $R_\alpha =1$. ($R_{\rm m}=200$
kills dynamo action for these parameters.) For comparison, in
Fig.~\ref{fig:wind2} we show a model with $r_3=1.4$, $R_\omega=20,
R_{\rm m}=150$, $R_\alpha =1$.  In all models, the magnetic pattern
is displaced perpendicular to the wind direction and a spectacular
magnetic tail can appear.

Of course, dynamo action is significant for the results obtained.
For comparison we performed a simulation with $R_\alpha = 0$ to
confirm that the magnetic field rapidly decays for such a flow.

We appreciate that a homogeneous wind velocity profile is certainly
a severe oversimplification, given the substantial density contrast
between the exterior medium and the galactic disc. Thus we implement
a wind profile that decays inside the galaxy. In order to mimic it
we multiply the originally uniform wind velocity by the factor $\exp
(-l/l_d)$ where $l$ is the distance a wind particle has traveled since
encountering the nominal edge of the galaxy at $r=r_w$ (with a
natural choice $r_w = 1$). To be specific, we choose $l_d = 4$.

This wind strongly reduces the efficiency of the dynamo and this
reduction grows with smaller $l_d$. Even for $l_d=4$ we have to use
$R_\alpha = 2$ and $R_\omega = 30$ to obtain a growing field instead
of $R_\alpha = 1$ and $R_\omega = 20$ for which magnetic field
becomes decaying.

Magnetic field evolution with an $x$-dependent wind velocity results in a
minor displacement of the magnetic structure from the galactic
centre, which is almost invisible in the plot
(Fig.~\ref{fig:wind3}), and the magnetic tail disappears.

The velocity field shown in Fig.~\ref{fig:wind3} has a non-vanishing
divergency. In principle, it can be argued that a $z$-component of
velocity would compensate this divergency. In the no-$z$ context,
such a component might be represented by an additional loss term.
However, to obtain some idea of the possible importance of this
omission, we added a term $v_y$, determined from the condition
$\partial v_x/\partial x +
\partial v_y / \partial y = 0$. The effect of this change was found
to be negligible.

\begin{figure}[ht!]
\includegraphics[height=0.335\textwidth]{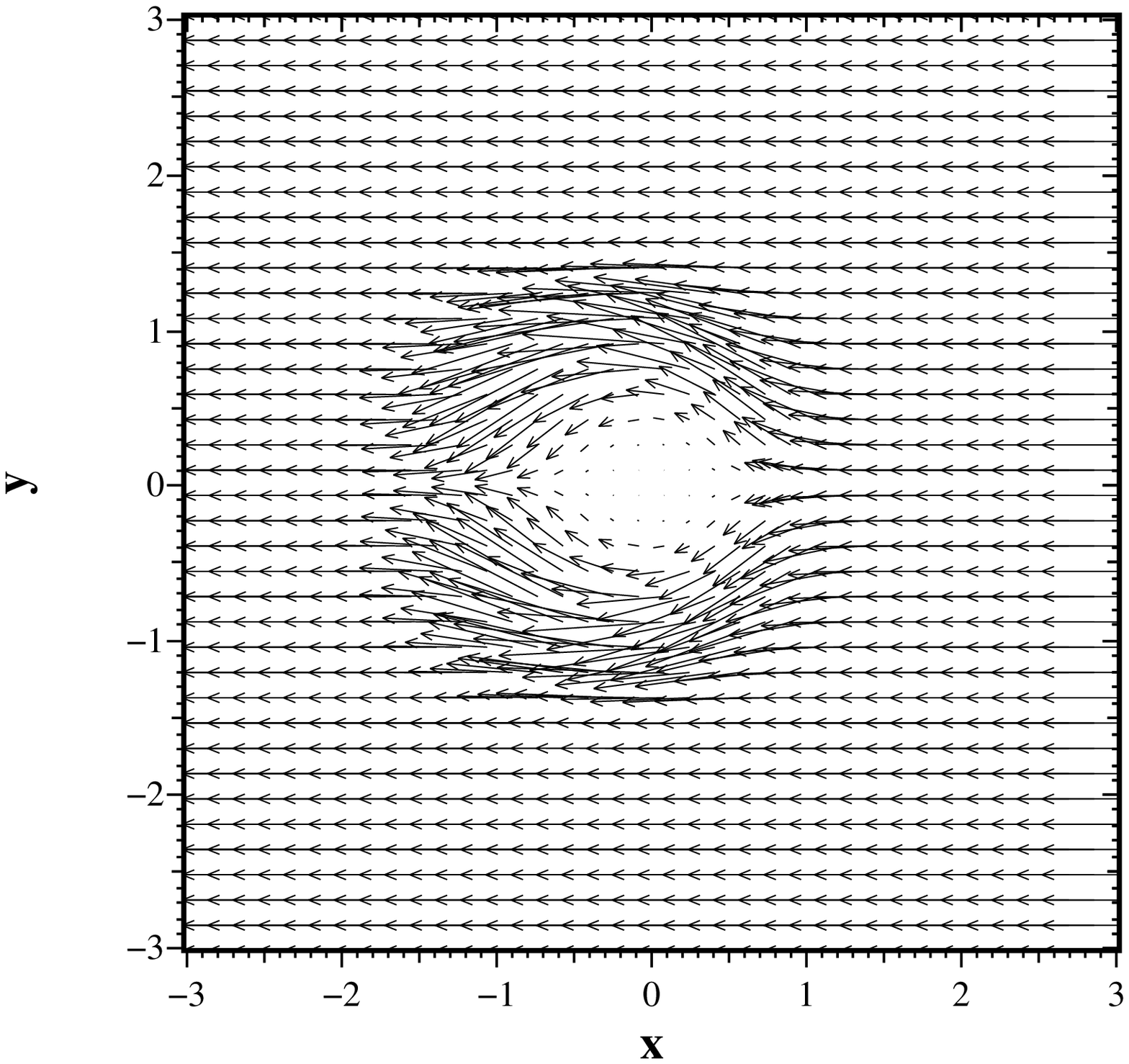}\\
\includegraphics[height=0.370\textwidth]{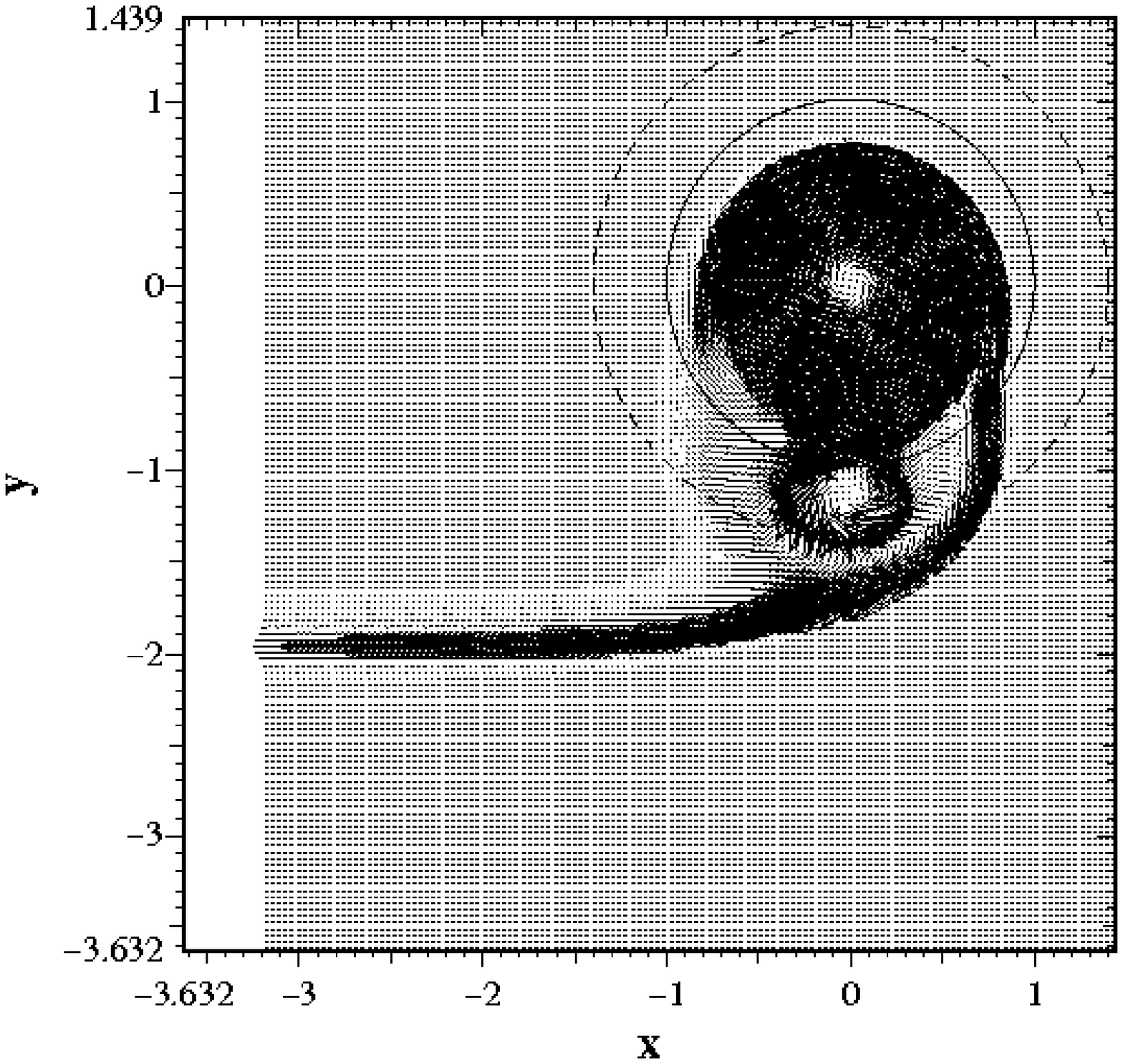}\\
\includegraphics[height=0.370\textwidth]{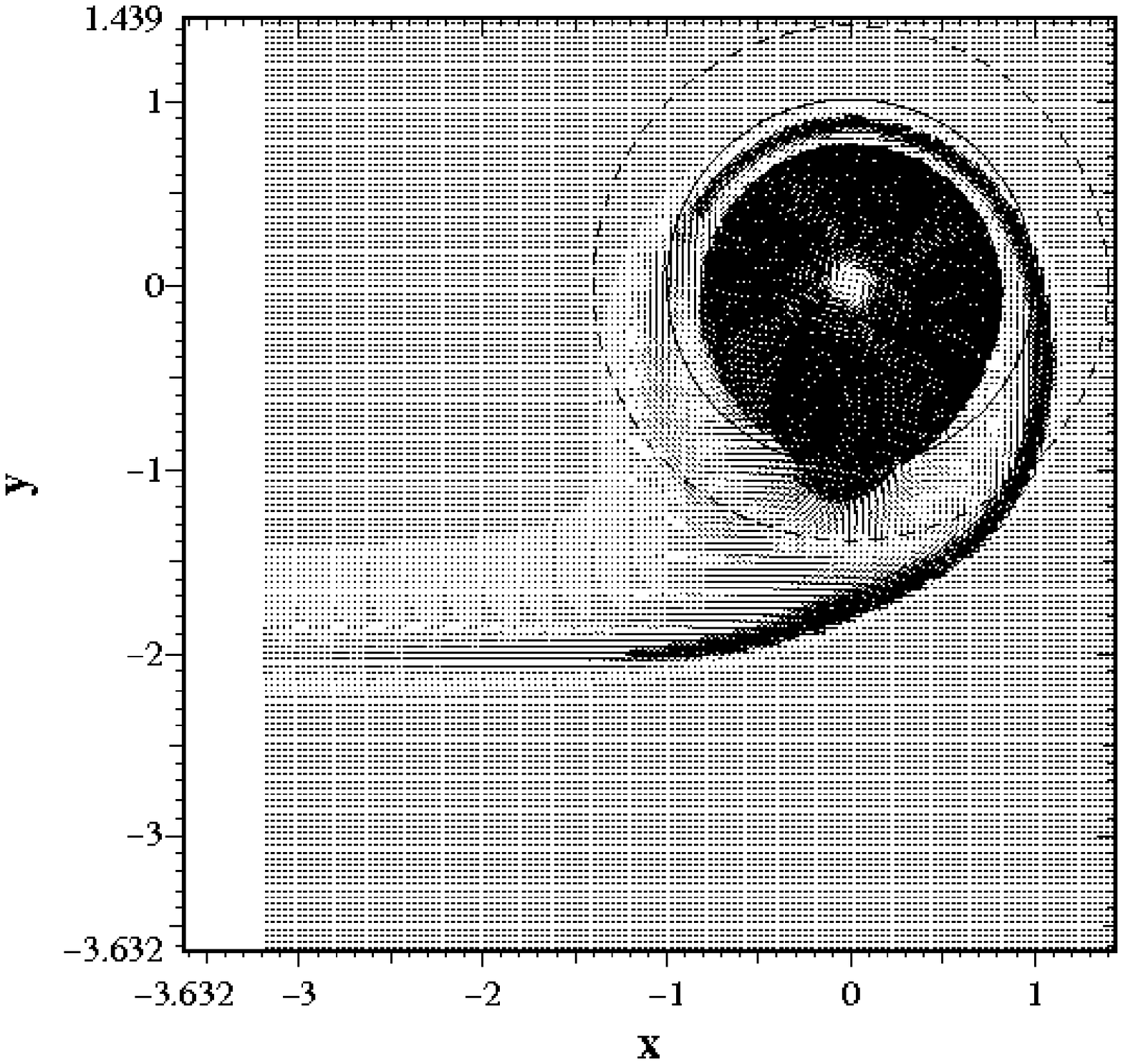}\\
\caption{Model with wind deflected around the galaxy ("draping
flow", top panel) and the corresponding magnetic structures.
Governing parameters are $R_\alpha=1, R_\omega=10, r_3=1.4$, with
$R_{\rm m}=100$ in the middle panel, and $R_{\rm m}=75$ in the
lower. Note that in the uppermost plot, the centre of the plot is
the centre of the galaxy.  }\label{fig:wind4}
\end{figure}

\begin{figure}
\includegraphics[height=0.370\textwidth]{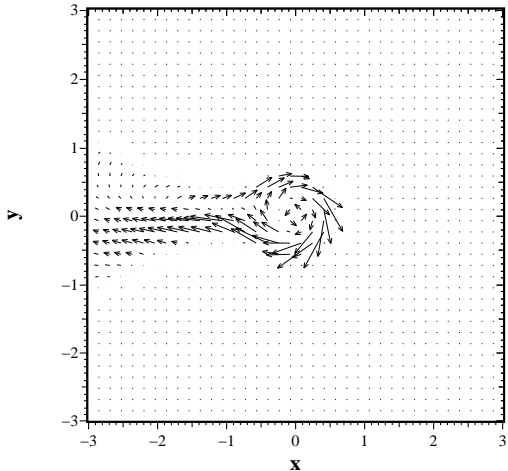}\\
\caption{Model with the draping flow interacting with a magnetic
configuration that is initially prescribed by dynamo action, but
then the dynamo action is switched off. A snapshot for just one
instant is presented. As in the other figures, the mean streaming
velocity is in the negative $x$-direction. } \label{fig:wind5}
\end{figure}

To explore an additional possibility, we follow a suggestion of
\cite{pd10} and also consider a wind that is deflected around the
galaxy, i.e. a case of a ``magnetic drape'' (Fig.~\ref{fig:wind4},
upper panel). Of course, we have to add extra terms involving $u_y$ to
Eqs.~(\ref{evolBr}), (\ref{evolBphi}).
An example of a stationary magnetic structure
(Fig.~\ref{fig:wind4}, middle panel) resulting from such a flow
shows again a general displacement of magnetic structure in the
direction perpendicular to the wind.

Details of the elongated magnetic structure are strongly model
dependent, as demonstrated in Fig.~\ref{fig:wind4}, for example,
compare the lower and middle panels ($R_{\rm m}=75$ and 100,
respectively).

Our modelling includes ongoing dynamo action, rather than
prescribing the initial magnetic structure, and allowing it to
evolve passively, subject to the ram pressure flow. We demonstrate a
distinction between two cases by considering a model with evolution
of an initially prescribed magnetic field which is affected by a
wind and is not supported by further dynamo action
(Fig.~\ref{fig:wind5}). Such a magnetic configuration decays and is
transported by the wind further and further away from the galaxy
centre. For cases with dynamo action, for some wind profiles the
magnetic configuration becomes a straight tail emerging from the
side of the galaxy in the direction downstream of the wind, while
the tail in the model shown in middle panel of Fig.~\ref{fig:wind4}
emerges on the upstream side, and is twisted around the galaxy. In
Fig.~\ref{fig:wind4}, lower panel, the tail originates on the left
side and is wound around the galaxy over about 270 degrees.  Note
that the end of the tail detaches from the galaxy and continues
downstream at the side of the galaxy where the wind decreases the
total gas velocity (negative $y$, ``south'' side). It is debatable
whether we should refer to a ``tail'' in these cases when describing
a ``wound around'' field, rather than considering this to be an
immediate result of the modified dynamo action within the galaxy.

\section{Discussion}
\label{disc}

We have presented the results of modelling of the effects of ram
pressure on the stationary, dynamo-generated, magnetic field
configuration in spiral galaxies. We
recognize a general displacement of the magnetic configuration from
the galaxy centre as well as, in some cases (not always, see
Fig.~\ref{fig:wind3}), magnetic tails propagating from the galaxy
into the surrounding medium. These effects are rather nontrivial and
have not been recognized in previous modelling attempts (Roediger
2009, Vollmer et al. 2009, 2012).

Figs.~\ref{fig:wind1} and \ref{fig:wind2} show that the wind
displaces the centre of the magnetic field pattern in a galaxy from
the nominal centre of mass. Surprisingly, this displacement is not in the
downstream direction of the wind flow, but perpendicular to this
direction. As the galaxy is assumed to rotate counterclockwise, the
wind increases the total gas velocity on the ``north'' side of the
galaxy and decreases it on the ``south'' side. It thus appears that
the large-scale field is displaced towards the galaxy in the region
where the gas velocity is reduced. The small-scale (turbulent)
magnetic field component, related to the star-formation activity in
a galaxy and hence the gas distribution, is not affected, in
agreement with the observation that the total radio emission is not
displaced.

If the ram speed of the wind is too large, it evacuates the
large-scale magnetic field from the galaxy and dynamo action is
killed. The critical velocity $U^*$ at which dynamo action is still
possible can be estimated as follows. Accepting for a crude estimate
that the maximal linear velocity of differential rotation $V =
\Omega R$ and taking into account the definitions of $R_\omega$ and
$R_m$ we obtain

\begin{equation}
U = V \, (R_{\rm m}/R_\omega) \, \lambda.
 \label{est}
\end{equation}
For $R_{\rm m} = 150$, $R_\omega = 20$ (close to the maximal wind
velocity for which dynamo still works) and $\lambda = 0.05$, $V =
250$\,km/sec we obtain $U^* \approx 100$\,km/sec.

Comparing the estimate $U^* = 100$\,km/sec with the observed
velocities of galaxies relative to the intracluster medium it is
necessary to take into account that the majority of our models do
not include the density contrast between the interstellar medium in
the galaxy (ISM) and the intracluster medium (ICM). Estimating the
limiting velocity of a galaxy $W$ in the cluster medium that is just
sufficient to kill dynamo action from the condition $\rho_{\rm ISM}
U^* = \rho_{\rm ICM} W$, and taking $\rho_{\rm ISM}/\rho_{\rm ICM}
\approx 10^{-2}$\,cm$^{-3} / 10^{-4}$\,cm$^{-3} \approx 100$ we
derive $W \approx 10^4$\,km/sec (here $\rho_{\rm ISM}$ and
$\rho_{\rm ICM}$ are typical ionized gas densities in the galactic
disc and the intracluster medium, respectively).

If $U > U^*$, the wind pushes the dynamo-generated magnetic field
out of the region where dynamo is active and hence kills the dynamo.
If $U \ll U^*$, the wind almost does not affect the dynamo-generated
magnetic field. If $U$ is slightly less than $U^*$, the main effect
is a less efficient dynamo action in the ``northmost'' part of the
galaxy where wind and rotation velocities reinforce. In contrast,
the magnetic field is larger in the  ``south'' part of the
galaxy where the velocities tend to cancel and dynamo action is
stronger. As a result, the magnetic configuration is shifted across
rather along the wind direction. There is thus a substantial
difference between the effect of a wind on a magnetic configuration
generated {\it in situ} (i.e. by dynamo action) from that on a
magnetic configuration generated before the wind impacts
on the galaxy. In our models the dynamo operates contemporaneously
with the wind, and so the wind
modifies the dynamo velocity field, and hence the dynamo properties.

The wind direction and the general displacement of the
main magnetic structure are mutually perpendicular. This must be into
account in determinations of galactic motion through the cluster
medium from observational data (cf. e.g. Pfrommer \& Dursi 2010).

The displacements obtained in the models are quite moderate. It is
impossible to push the centre of the magnetic configuration far
beyond the galaxy centre, or the boundary of the region occupied
by the bulk of the field far beyond its original position. This is
partly at least because in the absence of external dynamo sources
(here alpha-effect) the field rapidly decays. Note that the magnetic
vectors in the map of NGC~4535 cover the centre of the optical
structure of the galaxy, i.e. the displacement of magnetic structure
from the optical image is still quite modest.

Another unexpected result is the location of the magnetic tail: it
emerges from different regions of the disc depending on the details
of the model, and then propagates into the area where the wind
decreases the total velocity. The dynamo-generated field propagates
due to turbulent diffusion out of the area which rotates
significantly, and is caught by the wind, producing a tail.

Observations (We\.zgowiez et al. 2007) do not show the presence of
magnetic tails in  NGC~4535 and NGC~4501. On the other hand, the
spectacular magnetic tails presented in Fig.~\ref{fig:wind1}
may be related to the tail of polarized emission in NGC~4438
(Fig.~\ref{fig:n4438}; Vollmer et al. 2007, 2010). Our simple model
does not describe gas flows which might be associated with the
magnetic tails. It appears plausible that gas is also carried along
in the magnetic tail. Chung et al. (2007) found seven spiral
galaxies in the Virgo cluster with long, one-sided tails of neutral
gas.

The various features of magnetic field structures associated with
the ram pressure effects obtained in the models presented above have
different ranges of robustness. The magnetic structure is displaced
{\it perpendicularly} to the wind direction in all models. Magnetic
tails are quite ``fragile'' and disappear when the turbulent
magnetic diffusivity in the surrounding space is increased. The
value of turbulent magnetic diffusivity in galactic halos and
intracluster media is an important dynamo governing parameter, but
is difficult to estimate (e.g. Sokoloff \& Shukurov 1990). If we
include a deceleration of the wind in the disc, the tails are
smaller and its recognition becomes difficult -- see Sect.~3. The
tail disappears if an $x$-dependence is imposed on the wind velocity
(Sect.~3). Furthermore, the no-$z$ approximation can no longer be
expected to be valid far outside of the disc region, where {\it
inter alia} the estimate of $-B/h^2$ for the $z$-diffusion may in
turn be an over-estimate. Given that the dynamo does not operate in
the external region, the latter uncertainty can probably be formally
absorbed into the uncertainty concerning the external diffusivity.
Thus the presence of a magnetic tail cannot be considered a robust
feature of our simulations.

A statistical investigation of magnetic tails in cluster galaxies
could contribute to elucidation of the problem. Of course, further
details such as elongated magnetic structures and their particular
forms are even more uncertain and need further more detailed
modelling.

Our simulations are necessarily restricted to considering flows
parallel to the galactic plane. Real flows will be inclined at
arbitrary angles. We can split such a general flow into components
parallel to and perpendicular to the plane. We have separately
modelled the effects of a perpendicular flow (parallel to the
$z$-axis) in an {\it axisymmetric} dynamo model with dynamo effect
confined to the disc region. Here the effect is more clearly an
advection of the field structure in the direction of the flow,
effectively "lifting" the field from the galactic midplane. As the
field is moved out of the disc region, where the alpha-effect is
assumed to operate, in the absence of dynamo action the field
decays. The advection is thus naturally limited in extent. A
somewhat faster flow removes the field from the dynamo-active region
more rapidly than it can be regenerated, and so kills the dynamo
completely. Thus we feel our results are reasonably representative
of flows that are not too strongly inclined to the galactic plane.
In contrast, strong flows at large angles to the disc plane will
seriously weaken dynamo action.

Finally, we wish to stress again that the problem under discussion
is different from the classical problem of a magnetized wind
affecting a body with prescribed magnetic field (this problem in the
galactic context is addressed e.g. by Ruszkowski et al. 2012). The
interaction between the wind and the internal velocity field, and
hence the effect of the wind on dynamo action, cannot be avoided,
unless the internal magnetic field is completely shielded from the
external flow - which however is unrealistic. The problem is no
longer that of a dynamo with a simple circular flow. Hence, the
interpretation of an asymmetry in polarized radio intensity as a
naive effect of ram pressure is similarly unrealistic. A compression
of the internal field is possible only if the wind penetrates the
galaxy, which in turn must affect dynamo action. Clearly there are
plausible improvements that could be made to our no-$z$ model, but
we believe our results are generic. We have investigated a magnetic
field that is self-excited in a region that is directly affected by
the wind, i.e. the flow generating the dynamo is altered. This
problem has not attracted attention so far. In this preliminary
study we have attempted to isolate the basic physical effects. We
assume that the wind itself is non-magnetized and flows parallel to
the galactic plane, and use simple models to describe the wind
profile. These assumptions allowed us to use the simple no-$z$
approach. A more detailed modelling, including a magnetized wind
inclined to the galactic plane and extension to a 3D model, and /or
to go beyond a mean-field formulation of the problem, would be
instructive and should be investigated in future.

\section*{Acknowledgements}

We thank Marek We\.zgowiec for providing Figure~1 and Bernd Vollmer
for Figure~2. RB acknowledges support from the DFG Research Unit
FOR1254, and Marita Krause for a number of useful comments on the
text. An anonymous referee also helped to improve the paper. The
research was partially supported by the grant DFG--RFBR 08-02-92881.

\end{document}